# A Contour-Guided Deformable Image Registration Algorithm for Adaptive Radiotherapy

Xuejun Gu[1,2], Bin Dong[1,3], Jing Wang[2], John Yordy[2], Loren Mell[1], Xun Jia[1], and Steve B. Jiang[1]

[1]Center for Advanced Radiotherapy Technologies and Department of Radiation Medicine and Applied Sciences, University of California San Diego, La Jolla, CA 92037-0843, USA

[2]Department of Radiation Oncology, The University of Texas Southwestern Medical Center, Dallas TX, 75390 USA

[3]Department of Mathematics, The University of Arizona, Tucson, AZ, 85721-0089, USA

E-mails: xuejun.gu@utsouthwestern.edu (Gu) and sbjiang@ucsd.edu (Jiang)

In adaptive radiotherapy, deformable image registration is often conducted between the planning CT and treatment CT (or cone beam CT) to generate a deformation vector field (DVF) for dose accumulation and contour propagation. The auto-propagated contours on the treatment CT may contain relatively large errors, especially in low-contrast regions. A clinician's inspection and editing of the propagated contours are frequently needed. The edited contours are able to meet the clinical requirement for adaptive therapy; however, the DVF is still inaccurate and inconsistent with the edited contours. The purpose of this work is to develop a contour-guided deformable image registration (CG-DIR) algorithm to improve the accuracy and consistency of the DVF for adaptive radiotherapy. Incorporation of the edited contours into the registration algorithm is realized by regularizing the objective function of the original demons algorithm with a term of intensity matching between the delineated structures set pairs. The CG-DIR algorithm is implemented on computer graphics processing units (GPUs) by following the original GPU-based demons algorithm computation framework [Gu *et al*, Phys Med Biol. 55(1): 207-219, 2010]. The performance of CG-DIR is evaluated on five clinical head-and-neck and one pelvic cancer patient data. It is found that compared with the original demons, CG-DIR improves the accuracy and consistency of the DVF, while retaining similar high computational efficiency.

## 1. Introduction

Deformable image registration (DIR) is a critical component in modern image-guided online and off-line adaptive radiation therapy (Yan, 2008). It is a process to precisely establish voxel-to-voxel correspondence of two images collected at different times or with different imaging modalities (Hill *et al.*, 2001; Crum *et al.*, 2004). The established correspondence, namely the deformation vector field (DVF), has many applications in radiotherapy, such as facilitating auto-segmentation for anatomical changes (Brock *et al.*, 2005; Lu *et al.*, 2006; Rietzel and Chen, 2006; Chao *et al.*, 2008; Wang *et al.*, 2008; Xie *et al.*, 2008), estimating organ motion via 4D-CT images (Boldea *et al.*, 2008; Ehrhardt *et al.*, 2007; Yang *et al.*, 2008b; Zeng *et al.*, 2007), assisting the reconstruction of high-quality 4D-CT and CBCT images (Ren *et al.*, 2012; Wu *et al.*, 2011; Wang and Gu, 2013), and calculating accumulated dose (Yan *et al.*, 1999; Rietzel *et al.*, 2005; Keall *et al.*, 2005). All these applications rely on accurate DVFs generated by DIR algorithms.

During the past few decades, DIR has been studied in great detail and many algorithms have been proposed and developed (Holden, 2008; Kashani *et al.*, 2008; Crum *et al.*, 2004; Brock and Deformable Registration Accuracy Consortium, 2010). Among existing DIR algorithms, the image intensity based demons algorithm has been proven to be an efficient and robust algorithm (Pennec *et al.*, 1999; Wang *et al.*, 2005; Yang *et al.*, 2008a). In addition, high computational efficiency is achieved by implementing demons algorithms on GPU (Gu *et al.*, 2010). In the demons algorithm, the force used to deform an image is proportional to the gradients of moving and/or target image intensity. Therefore, the accuracy of estimating the image intensity gradient has great influence on the accuracy of the image deformation. As a result, deforming low-contrast and/or high-noise images with high accuracy is challenging using the demons algorithm (Nithiananthan *et al.*, 2009; Yang *et al.*, 2008b; Zhong *et al.*, 2010).

In adaptive radiotherapy, DVFs generated in DIR algorithms are utilized mainly for dose accumulation and auto-contour propagation (Thor *et al.*, 2011). As uncertainties and errors exist in DVFs, the auto-propagated contours are apt to be distorted and hence required to be inspected and edited by physicians before their usage in treatment re-planning (Zhang *et al.*, 2007; Wang *et al.*, 2008; Reed *et al.*, 2009). Physicians' inputs correct the errors in contours generated by DVF, but not the errors in the DVF itself. Calculated dose accumulation will be inaccurate due to these errors, if the DVF is inconsistent with the edited contours. In this paper, we propose to use edited contours to guide another run of DIR to reduce errors in DVF and to increase the consistency between DVF and final contours.

Studies using contours to guide image registrations have been conducted for many years. For example, Shih *et al.* (Shih *et al.*, 1997) proposed an automated contour-model-guided DIR model to register a pair of MRI images. In the first step of their algorithm, authors established the geometric correspondence between contours in moving and reference MRI images. Then, based on the established correspondence, a non-linear transformation was determined using a weighted local coordinate system to deform the moving image. This algorithm is a pure geometrical-based image registration and the accuracy decreases with the distance away from contour points. Later, Lie and Chuang (Lie and Chuang, 2003) designed a hybrid algorithm for deforming thermographs. They developed an algorithm combining the scheme of local contour matching and the method of global surface-spline fitting of control points. The registration accuracy achieved by their algorithm relies on the density of control points, which may impose a very heavy computational burden when dense control points are used to achieve high registration accuracy.

In this paper, we propose a GPU-based contour-guided DIR (CG-DIR) algorithm, which honors global accuracy of registration while retaining high computational efficiency. The CG-DIR model is a derivation of the demons algorithm, where the objective function of demons is revised, but the optimization scheme is unchanged. The objective function of the original demons has a term of summation of image intensity difference and a term of diffusion regularization of DVFs. In the proposed model, we add an additional regularization term using the regions of interest (ROIs) manifested by physicians' edited contours pairs. The optimization scheme is still the original demons iterative method, where a step of DVF updating and a step of DVF smoothing are performed during each iteration. With the preservation of the original GPU-based demons computational framework, GPU-based CG-DIR algorithm is able to achieve high computational efficiency. In this paper, we first detail the proposed CG-DIR model and its algorithm. Then, the performance of the new algorithm is quantitatively and qualitatively evaluated on five head-and-neck cancer patient data sets and one pelvic patient data set.

## 2. Methods and Materials

*2.1 Contour-guided DIR model and algorithm*

The goal of demons registration is to have the intensity matching between static and moving image through deforming moving image. The demons algorithm can be cast into a generalized optimization framework (Vercauteren *et al.*, 2009):

$$E(\mathbf{u}) = \frac{1}{2} \|I_0 \circ \mathbf{u} - T_0\|^2 + \frac{\alpha}{2} \|\nabla \mathbf{u}\|^2, \quad (1)$$
$$\mathbf{u} = \arg\min E(\mathbf{u})$$

Here, an optimal deformation vector field $\mathbf{u}$ is estimated by minimizing an energy function between a moving image $I_0$ and a target image $T_0$. $\alpha$ is a regularization parameter for controlling the smoothness of $\mathbf{u}$ over the image space.

In adaptive radiotherapy, $I_0$ often represents a planning CT (pCT) image and $T_0$ refers to a treatment CT (tCT) image. With physician delineated contours in pCT and edited contours in tCT, a constraint term defined by contours can be incorporated into equation (1) as:

$$E(\mathbf{u}) = \frac{1}{2} \|I_0 \circ \mathbf{u} - T_0\|^2 + \sum_{i=1}^{k} \frac{\lambda_i}{2} \|I_i \circ \mathbf{u} - T_i\|^2 + \frac{\alpha}{2} \|\nabla \mathbf{u}\|^2, \quad (2)$$

$$\mathbf{u} = \arg\min E(\mathbf{u})$$

Here, $I_i$ and $T_i$ denote modified images constructed by incorporating the *i*th contour pair set on pCT and tCT into original images $I_0$ and $T_0$, respectively; and $k$ is the total number of modified contour pairs. The image modification is achieved with $I_i = \left(1 + \frac{M_i(\mathbf{x})}{2}\right) I_0$ and $T_i = \left(1 + \frac{N_i(\mathbf{x})}{2}\right) T_0$, where $M_i(\mathbf{x}) = \begin{cases} \pm 1 & \mathbf{x} \in O_i^p \\ 0 & else \end{cases}$ and $N_i(\mathbf{x}) = \begin{cases} \pm 1 & \mathbf{x} \in O_i^t \\ 0 & else \end{cases}$. $O_i^p$ (or $O_i^t$) represents a spatial domain that manifests a volume enclosed by the *i*th contour on pCT (or tCT). We refer $O_i^p$ and $O_i^t$ as regions of interest (ROIs) and $I_i$ and $T_i$ as modified images (MIs). $M_i(\mathbf{x})$ and $N_i(\mathbf{x})$ are positive when ROIs' intensity is higher than its surrounding and negative when lower. For simplification, $\lambda_i$ is a regularization parameter for the *i*th contour pair set, which controls the influence of ROIs registration on the entire images registration.

Similar to the alternative optimization strategy proposed in Vercauteren *et al.* (Vercauteren *et al.*, 2009), an auxiliary variable **c** is introduced to equation (2) to enable an relaxation minimization:

$$E(\mathbf{c}, \mathbf{u}) = \frac{1}{2} \|I_0 \circ \mathbf{c} - T_0\|^2 + \sum_{i=1}^{k} \frac{\lambda_i}{2} \|I_i \circ \mathbf{c} - T_i\|^2 + \frac{\beta}{2} \|\mathbf{c} - \mathbf{u}\|^2 + \frac{\alpha}{2} \|\nabla \mathbf{u}\|^2, \quad (3)$$

$$(\mathbf{c}, \mathbf{u}) = \arg\min E(\mathbf{c}, \mathbf{u})$$

The algorithm is minimizing **c** and **u** in alternative steps. The first step starts with a given **u** and optimizes $\text{argmin}_{\mathbf{c}} E_1(\mathbf{c}) = \frac{1}{2} \|I_0 \circ \mathbf{c} - T_0\|^2 + \sum_{i=1}^{k} \frac{\lambda_i}{2} \|I_i \circ \mathbf{c} - T_i\|^2 + \frac{\beta}{2} \|\mathbf{c} - \mathbf{u}\|^2$, where optimization make steps from $\mathbf{c} = \mathbf{u}$. The second step solves the regularization by optimizing $\text{argmin}_{\mathbf{u}} E_2(\mathbf{u}) = \frac{\alpha}{2} \|\nabla \mathbf{u}\|^2 + \frac{\beta}{2} \|\mathbf{c} - \mathbf{u}\|^2$ with given **c**. In the second step, variety of regularizations have been proposed (Cachier and Ayache, 2004). Here, we choose an optimal regularization reached by the convolution of the deformation field **u** with a Gaussian smoothing kernel $\mathbf{G}(\alpha)$ (Vercauteren *et al.*, 2009).

In the first step, given the current deformation **u**, an updated (or displacement) deformation vector d**u** can be calculated by minimizing:

$$\arg\min_{d\mathbf{u}} E_1(d\mathbf{u}) = \frac{1}{2} \|I_0 \circ \mathbf{u} \circ (\mathbf{Id} + d\mathbf{u}) - T_0\|^2 + \sum_{i=1}^{k} \frac{\lambda_i}{2} \|I_i \circ \mathbf{u} \circ (\mathbf{Id} + d\mathbf{u}) - T_i\|^2 + \frac{\beta}{2} \|d\mathbf{u}\|^2, \quad (4)$$

where $I_0 \circ \mathbf{u}$ and $I_i \circ \mathbf{u}$ are deformed images with a given deformation **u**. For simplification, $I_0 \circ \mathbf{u}$ is denoted as $\widetilde{I_0}$ and $I_i \circ \mathbf{u}$ as $\widetilde{I_i}$. With a small displacement d**u**, we can approximate the moving image term with a Taylor expansion: $I_0 \circ \mathbf{u} \circ (\mathbf{Id} + d\mathbf{u}) =$ $\widetilde{I_0} \circ (\mathbf{Id} + \mathrm{d}\mathbf{u}) \approx \widetilde{I_0} + \nabla\widetilde{I_0} \cdot \mathrm{d}\mathbf{u}$ and $I_i \circ \mathbf{u} \circ (\mathbf{Id} + \mathrm{d}\mathbf{u}) = \widetilde{I_i} \circ (\mathbf{Id} + \mathrm{d}\mathbf{u}) \approx \widetilde{I_i} + \nabla\widetilde{I_i} \cdot \mathrm{d}\mathbf{u}$.
Then, equation (4) becomes:

$$\arg\min_{\mathrm{d}\mathbf{u}} E_1(\mathrm{d}\mathbf{u}) \approx \frac{1}{2} \left\| \widetilde{I_0} + \nabla\widetilde{I_0} \cdot \mathrm{d}\mathbf{u} - T_0 \right\|^2 + \sum_{i=1}^{k} \frac{\lambda_i}{2} \left\| \widetilde{I_i} + \nabla\widetilde{I_i} \cdot \mathrm{d}\mathbf{u} - T_i \right\|^2 + \frac{\beta}{2} \|\mathrm{d}\mathbf{u}\|^2 \quad (5)$$

In equation (5), we rewrite $\mathrm{d}\mathbf{u} = v\mathbf{n}$, where, $v$ and $\mathbf{n}$ are the amplitude and the moving direction of the displacement vector $\mathrm{d}\mathbf{u}$, respectively. Assuming that the displacement is determined on the projection $\mathrm{d}\mathbf{u}$ onto $\mathbf{n} = \frac{\nabla\widetilde{I_0}}{|\nabla\widetilde{I_0}|}$ (Cachier *et al.*, 1999), the optimization problem defined in equation (5) can be simplified from $\mathrm{d}\mathbf{u} = \arg\min E(\mathrm{d}\mathbf{u})$ to $v = \arg\min E(v)$. Now the equation (5) becomes:

$$\arg\min_{v} E_1(v) \approx \frac{1}{2} \left\| (\nabla\widetilde{I_0} \cdot \mathbf{n})v - (T_0 - \widetilde{I_0}) \right\|^2 + \sum_{i=1}^{k} \frac{\lambda_i}{2} \left\| (\nabla\widetilde{I_i} \cdot \mathbf{n})v - (T_i - \widetilde{I_i}) \right\|^2 + \frac{\beta v^2}{2} \quad (6)$$

Applying the first-order optimality condition $\frac{\partial E_1(v)}{\partial v} = 0$, we obtain a solution of equation (6):

$$v = \frac{(\nabla\widetilde{I_0} \cdot \mathbf{n})(T_0 - \widetilde{I_0}) + \sum_{i=1}^{k} \lambda_i (\nabla\widetilde{I_i} \cdot \mathbf{n})(T_i - \widetilde{I_i})}{(\nabla\widetilde{I_0} \cdot \mathbf{n})^2 + \sum_{i=1}^{k} \lambda_i (\nabla\widetilde{I_i} \cdot \mathbf{n})^2 + \beta}; \quad (7)$$

where $\beta$ can be defined a spatial variable such as: $\beta = (T_0 - \widetilde{I_0})^2$ (Thirion, 1998).

*2.2 Implementation of CG-DIR on GPU*

As described in Section 2.1, CG-DIR is derived using a step of Taylor expansion and truncation, which implies that we can only allow a small and local displacement in registration. In order to reduce the magnitude of the displacement with respect to discretized image voxel size, a multi-scale strategy is adopted in CG-DIR algorithm. Registration iterations start with the lowest resolution, then the moving vectors obtained at a coarser level are up-sampled to serve as an initial solution at the next finer level. In this study, the entire registration is accomplished with two scales. The original images are only down-sampled once by a factor of 2 in each dimension. It was found that further down-sampling does not help improving either registration accuracy or efficiency. At coarse resolution levels, demons registration is adopted to generate an approximate DVF. At the finest level, the moving vectors are updated by the proposed CG-DIR algorithm. One major reason of adopting demons at coarse resolution level is to reduce computational burden. Also, the moving vectors obtained with demons at coarse resolution level are accurate enough to serve as an initial guess for the fine level CG-DIR registration.

Since the ROIs delineated by physicians are considered as the focus of our registration, the stopping criterion in CG-DIR is designed to place emphasis on the registration accuracy of these ROIs. Each ROI is expanded with 1cm margin. The union of the expanded ROIs, referred to as UROI in this paper, is the designed region to estimate the stopping metric. We defined a root mean square error of intensity difference (RMS-ID) at the *k*th iteration in UROI as: $\delta^k = \sqrt{1/N \sum (I^k(\mathbf{x}) - T(\mathbf{x}))^2}$, with $\mathbf{x} \in \text{UROI}$. Then, a stopping metric is chosen as the change of RMS-ID in successive iterations: $l^{(k)} = \delta^k - \delta^{k-1}$, and use $l^{(k-10)} - l^{(k)} \leq \varepsilon$, where $\varepsilon = 1.0 \times 10^{-4}$, as our stopping criterion.

In the CG-DIR algorithm, there are three regularization parameters involved: α, β, and $\lambda_i$. These three regularization parameters have different functionality in the proposed CG-DIR algorithm. α, corresponding to the standard deviation of a Gaussian filter in our implementation, ensures a spatial smoothness of the deformation vector **u**. In our study, we empirically choose a unit standard deviation for the Gaussian filter. β, related to image noise, is estimated locally as: $\beta = (T_0 - \widetilde{I_0})^2$ (Thirion, 1998; Vercauteren *et al.*, 2009). $\lambda_i$ is a parameter designed to balance the registration efforts between the modified image pair and the original image pair. The value of $\lambda_i$ depends on the intensity ratio between modified image ROIs and original image ROIs. In this work, the value of $\lambda_i$ is found by a trial-and-error strategy and $\lambda_i \sim 1.0 \times 10^{-3}$ gives good deformation results for all cases studied.

The proposed CG-DIR algorithm was derived based on the original GPU-based demons algorithm (Gu *et al.*, 2010) and implemented on the NVIDIA CUDA enabled GPU platform. We adopt the following data parallel GPU kernels including: kernel 1. ***a low-pass filter kernel*** to smooth images and deformation vectors; kernel 2. ***a gradient kernel*** to calculate the gradient of images; kernel 3. ***a moving vector kernel*** to calculate displacement vectors and update deformation vectors; kernel 4. ***an interpolation kernel*** to deform images with moving vectors and update moving vectors; kernel 5. ***a comparison kernel*** to calculate the stopping criteria to terminate iterations properly. Here, the low-pass filter kernel is accomplished by convolving the image with a three-dimensional Gaussian filter of a unit standard deviation. Taking advantage of the separability property of the Gaussian kernel, the convolution is implemented in $x$, $y$, and $z$ directions sequentially.

The CG-DIR algorithm is summarized as follows:

| **Algorithm A:** |
|---|
| **Initialization:** |
|   Set initial DVF $\mathbf{u_0} := 0$; |
|   Down-sample images ($I_0$, $T_0$) to the coarsest resolution; |
| **A. Coarse-level registration:** |
|   Perform demons registration between down-sampled $I_0$ and $T_0$ images; |
|   Up-sample moving vector **u** to a finer resolution level; |
| **B. Finest-level registration:** |
|    Create modified images $I_i$ and $T_i$; |

**loop** {*over n until convergence*}
1. Given the DVF at the *n*-th iteration $\mathbf{u^n}$, compute the displacement vector: $d\mathbf{u}^{n+1} = v^{n+1}\mathbf{n}^{n+1}$, where $v^{n+1}$ is from eq. (7) and $\mathbf{n}^{n+1} = \frac{\nabla I_0^n}{|\nabla I_0^n|}$ (kernels 2 and 3);
2. Smooth the displacement vector $d\mathbf{u^{n+1}} \leftarrow d\mathbf{u^{n+1}} \otimes \mathbf{G}$, here $\mathbf{G}$ is a Gaussian smoothing operator (kernel 1);
3. Update DVFs with a composite addition(Vercauteren *et al.*, 2009) $\mathbf{u^{n+1}} = \mathbf{u^n} \circ (\mathbf{Id} + d\mathbf{u^{n+1}})$ (kernel 4), here **I** is an identity matrix;
4. Smooth the updated DVF $\mathbf{u^{n+1}} = \mathbf{u^{n+1}} \otimes \mathbf{G}$ (kernel 1);
5. Warp images $I_0 \circ \mathbf{u^{n+1}}$ and $I_i \circ \mathbf{u^{n+1}}$ (kernel 4);
6. Calculate $l^{(n)}$ to assess the stopping metric $l^{(n-10)} - l^{(n)} \leq \varepsilon$ (kernel 5).

e**nd loop**

*2.3 Evaluation*

The performance of CG-DIR is evaluated with clinical pCT and tCT image pairs of five head-and-neck cancer patients and one pelvis cancer patient. The tCT images were acquired 3-4 weeks after the first fraction of treatment due to the significant changes in anatomy. In this study, both pCT and tCT images are segmented to exclude couch, masks, and other accessories. All DIR analysis is conducted on the segmented images. The original pCT and tCT images have dimensions around ~512×512×150. Due to the limitation of GPU memory, both tCT and pCT are down-sampled in axial planes and cropped in three dimensions. The images in this study have a resolution of 1.97×1.97×2.5 mm$^3$. Representative organs with low contrast image intensity to background such as GTV, parotids, rectum, and bladder were selected for testing and evaluating the CG-DIR.

Direct evaluation of DVF accuracy is difficult in clinical cases due to the lack of ground-truth DVFs. However, other information, such as image intensity, structure volume, and contours, is often available. Hence, indirect evaluation approaches through the usage of available ground-truth information is commonly adopted. More specifically, the calculated DVF is used to deform the image and ROI volumes, or to propagate the contours. Then, the accuracy of DVF can be estimated in term of difference between the deformed and the target image intensities, overlapping rate of the deformed and the target ROI volumes, or distance between the propagated contours and the target contours,.

In this study, the performance of the CG-DIR algorithm is first evaluated using image intensity difference. Two image intensity based assessment metrics are adopted: root mean square of intensity difference (RMS-ID) and correlation coefficient (CC) between the deformed pCT and the tCT. In the study, we calculate RMS-ID and CC over the entire image as well as the confined UROI.

The accuracy of the resulting DVF from the CG-DIR algorithm is further assessed using Dice Similarity Coefficient (DSC). DSC assessment of volume overlapping, is a widely used similarity metric for evaluating deformable image registration results (Zhang *et al.*, 2007; Wang *et al.*, 2008; Xie *et al.*, 2008). DSC is defined as: $DSC = \frac{2(V_D \cap V_T)}{(V_D + V_T)}$, where $V_D$ is the deformed volume of an ROI and $V_T$ is the target volume of the same ROI. In this study $V_T$ is the ROI volume enclosed by clinician manually edited contours. DCS measures the volume overlap of $V_D$ and $V_T$, zero indicating no overlapping and one indicating an exact volume match. We apply this metric to solid organs like GTV and parotids.

The third type of comparison is contour matching, where the evaluated contours were propagated from pCT to tCT using the calculated DVF. Since DVFs generated by the CG-DIR algorithm only establish a 3D voxel-to-voxel correspondence map between pCT and tCT, when 2D axial-plane defined contour points on pCT are mapped onto rCT, points can easily cross the plane. Such a point-to-point mapping will destroy points order in contour definition and generate non-smooth contours. Extra efforts are needed to reorganize points and smooth contours. To avoid those extra tasks, we adopt a volume deformation strategy (Wang *et al.*, 2008): 1) deforming ROI volumes; 2) using morphological "opening" operation to remove small islands; 3) extracting boundary to obtain the propagated contours.

Quantitatively, the contours comparison is assessed with surface-to-surface distance (SSD), which is defined as the shortest Euclidean distance of the deformed points on contours to the target surface. In this study, the deformed contours of bladder and rectum are evaluated with this metric. The main reason is that the accurate deformation of organ walls rather than the fillers for these two organs are more critical in therapy treatment. In particular, we choose several basic statistical quantities of SSD to quantify the bladder and rectum deformation in a pelvic cancer case: 1) mean value of SSD (M-SSD) measuring how close the deformed surface matched the target surface; 2) standard deviation of SSD values (SD-SSD) quantifying how deformed points scattered around M-SSD; and 3) 95 percentile SSD value gauging peak SSD values.

**3. Experimental Results**

The performance of CG-DIR algorithm is first assessed in term of image intensity on five head-neck cancer patient cases and one pelvic cancer patient case. Figure 1 illustrates intensity difference images of an example case (Case HN 5). Before deformable image registration, a large discrepancy of image intensity is observed, as shown in Figure 1(a). As the moving image is deformed by either demons algorithm or CG-DIR algorithm, the intensity inconsistency is considerably reduced, as shown in Figure 1(b) and (c). By comparing Figure 1(b) and (c), we can tell that these two difference images are very similar. We further illustrate intensity difference between CG-DIR deformed image and demons deformed image in Figure 1(d). Quantitative evaluation of intensity difference of entire and regional registered image pairs is listed in Table 1. In this study, regional evaluation volume is chosen as UROI defined in Section 2.2.2. RMS-ID and CC values are very similar across these two algorithms, only slight improvement inside UROIs are observed by using CG-DIR algorithm in three out of 6 cases (HN 3, HN 4, and HN 5). It is not surprising that intensity based metrics cannot differentiate these two algorithms. The CG-DIR algorithm focuses on improving DVF accuracy regionally near delineated structure(s) but does not disturb deformation of the entire image. Thus, close values of RMS-ID and CC of entire images between two algorithms are consistent with the intention of algorithm design. Due to low contrast in UROIs, the accuracy improvement of DVFs will not have significant impact on image intensity maps, thus, intensity-based metrics, RMS-ID and CC, are not sensitive to the improvements yielded by the CG-DIR algorithm.

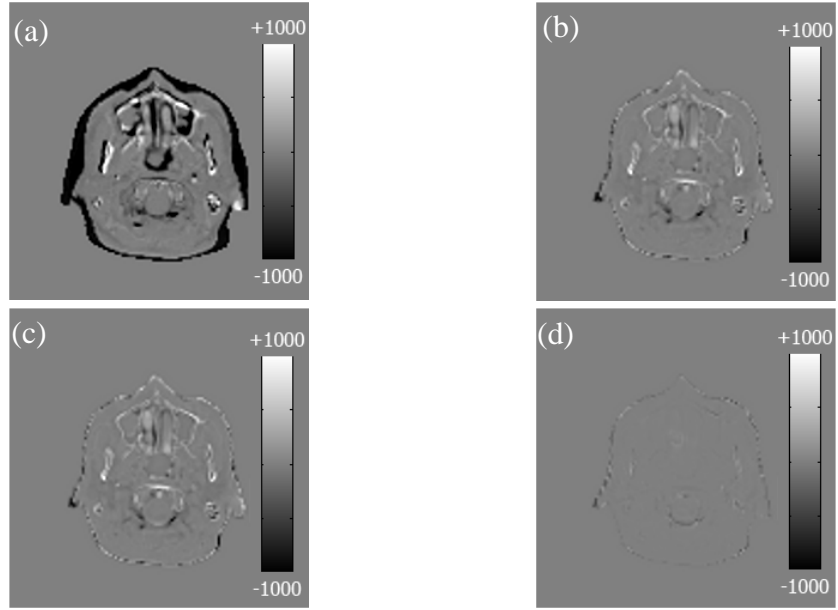

**Figure 1.** An example (Case HN 5) of difference image (scale in [-1000 1000] HU) at an axial cross section (a) between moving image and target image; (b) between demons deformed image and target image; (c) between CG-DIR deformed image and target image; and (d) between CG-DIR deformed image and demons deformed image.

**Table 1.** Root mean square of intensity difference and correlation coefficients of registered image pairs for six examined cases over entire image region (All) and in UROI.

|  |  | RMS-ID (HU) | | CC | |
|---|---|---|---|---|---|
|  |  | Demons | CG-DIR | Demons | CG-DIR |
| HN 1 | All | 166.84 | 174.17 | 0.961 | 0.960 |
|  | UROI | 2.58 | 2.82 | 0.927 | 0.923 |
| HN 2 | All | 68.00 | 68.36 | 0.994 | 0.990 |
|  | UROI | 1.12 | 1.17 | 0.981 | 0.980 |
| HN 3 | All | 82.01 | 81.45 | 0.987 | 0.987 |
|  | UROI | 0.73 | 0.70 | 0.978 | 0.979 |
| HN 4 | All | 74.87 | 72.97 | 0.980 | 0.983 |
|  | UROI | 1.28 | 1.23 | 0.960 | 0.963 |
| HN 5 | All | 115.43 | 111.31 | 0.976 | 0.977 |
| | UROI | 0.31 | 0.30 | 0.947 | 0.951 |
| Pelvic | All | 75.17 | 77.90 | 0.990 | 0.989 |
| | UROI | 0.23 | 0.23 | 0.923 | 0.923 |

Figure 2 shows an example result (Case HN 4), where manual drawn contours and auto-propagated contours are overlaid on tCT images. As we can see, big discrepancy exists between contours manually redrawn on tCT by physicians (blue lines) and those pCT contours mapped through rigid registration (red lines). This discrepancy is reduced when pCT contours is mapped through demons deformable registration (yellow lines). Further discrepancy reduction is achieved by mapping pCT contours through CG-DIR (green lines). This trend is quantitatively verified by examining the DSCs listed in Table 2. In our DSC evaluation, $V_T$ is physicians delineated structure volumes on tCT and $V_D$ is mapped (or deformed) structure volumes from pCT. As listed in Table 2, by applying rigid registration to map contours, the mean DSC value for evaluated structures varies from 0.02 ~ 0.40 (average, 0.25). The values of DSC increase to 0.7~0.9 when $V_D$ is demons deformed ROI volume. Further increasing of DSC to ~0.98 was observed when deforming ROI volume with CG-DIR algorithm.

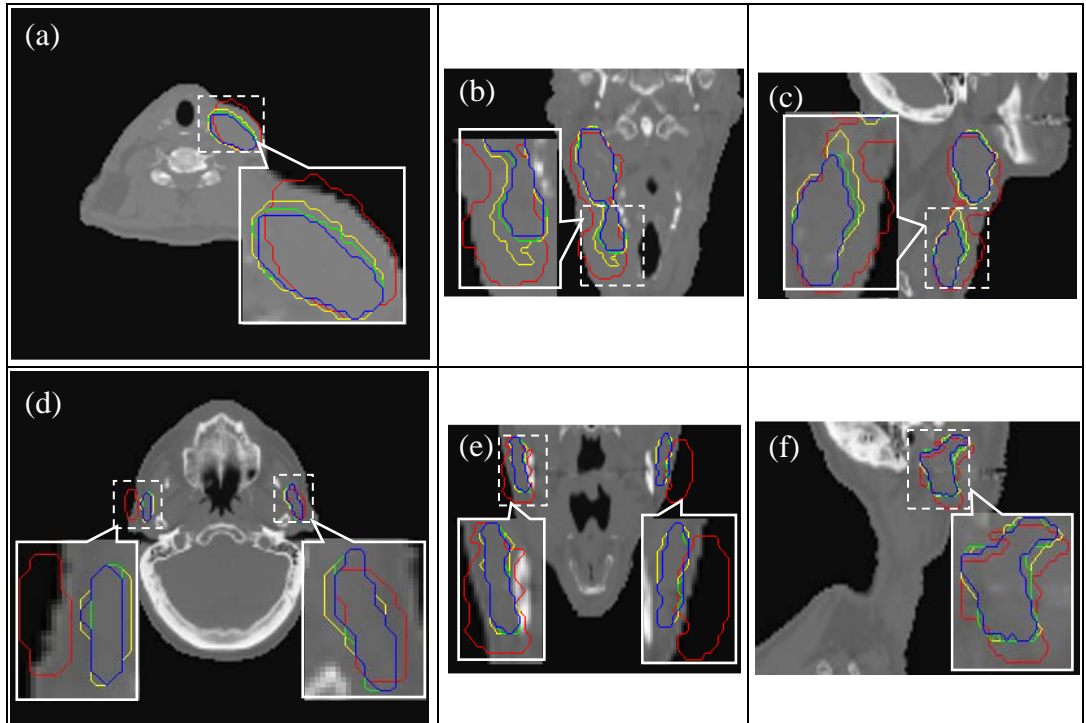

**Figure 2.** GTV and parotid contours overlaid on axial tCT (a and d, respectively), coronal (b and e respectively), sagittal (c and f, respectively) images for case HN4. Red lines: manual contours on pCT and mapped on tCT through rigid registration. Blue lines: manually edited contours on tCT. Yellow line: deformed pCT contours through demons registration. Green line: deformed pCT contours through CG-DIR registration.

**Table 2.** Dice similarity coefficients between physicians' manually delineated parotids and GTV volumes and mapped volumes using different registration algorithms.

|  |  | Rigid Registration | Demons | CG-DIR |
|---|---|---|---|---|
| HN 1 | Left Parotid | 0.03 | 0.70 | 0.94 |
|  | Right Parotid | 0.02 | 0.92 | 0.97 |
| HN 2 | Left Parotid | 0.39 | 0.94 | 0.98 |
|  | Right Parotid | 0.42 | 0.90 | 0.97 |
| HN 3 | Left Parotid | 0.33 | 0.94 | 0.99 |
|  | Right Parotid | 0.30 | 0.95 | 0.99 |
| HN 4 | Left Parotid | 0.27 | 0.90 | 0.96 |
|  | Right Parotid | 0.07 | 0.94 | 0.98 |
|  | GTV | 0.32 | 0.89 | 0.97 |
| HN 5 | Left Parotid | 0.29 | 0.89 | 0.95 |
|  | Right Parotid | 0.24 | 0.94 | 0.98 |
|  | GTV | 0.40 | 0.95 | 0.99 |

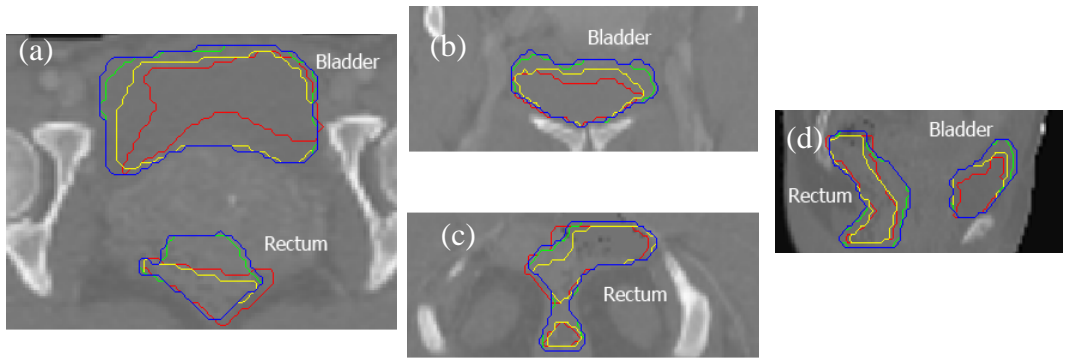

**Figure 3.** The bladder and rectum contours overlaid on tCT axial (a), coronal (b and c),, and sagittal (d) images for the pelvic case. Coronal cross-section for bladder and rectum contours are shown in b and c, respectively. Red lines: manual contours on pCT and mapped on tCT through rigid registration. Blue lines: manual contours on tCT. Yellow line: deformed contours through demons registration. Green line: deformed contours through CG-DIR registration.

Further evaluation of the CG-DIR algorithm is conducted on a pelvic cancer patient case using SSD. Figure 3 illustrates mapped and manually drawn contours overlaid on tCT images. From visual inspection, the CG-DIR mapped contours have the best agreement with manual tCT contours, and the rigid mapping has the worst one. Quantitative measurement of contours agreement is given by calculating SSD. Table 3 lists some basic statistical values of SSD, including mean, standard deviation, and 95 percentile. As shown in Table 3, 95 percentile errors of CG-DIR wrapped bladder and rectum surface points have errors around 2~3mm, which indicates that only 5% points on wrapped surfaces have errors larger than a voxel size. In contrast, demons wrapped surfaces have a relative large error, where 95% percentile errors are 3~4 voxel size.

**Table 3.** Statistical function of SSD between clinician manually delineated rectum and bladder contours and mapped contours using different registration algorithms

| SSD (mm) | Demons | | | CG-DIR | | |
|---|---|---|---|---|---|---|
| | Mean | Standard Deviation | 95 percentile | Mean | Standard Deviation | 95 percentile |
| **Bladder** | 2.75 | 5.80 | 7.81 | 0.74 | 1.09 | 3.09 |
| **Rectum** | 2.44 | 2.66 | 8.29 | 0.48 | 0.74 | 1.95 |

As mentioned above, the CG-DIR algorithm is derived from the demons algorithm and its implementation follows the original demons computational framework. (Gu *et al.*, 2010). Compared to the original demons algorithm, the CG-DIR algorithm has three extra computational components at each iteration for vector field updating: 1) enhanced image gradient calculation; 2) enhanced image deformation; 3) composite moving vector calculation using equation (7). Due to these extra computational components, the CG-DIR algorithm is demanding on graphic memory, because additional variables are needed to store enhanced images, their gradient fields, as well as a binary UROI map used in calculating stopping criteria. The performance of the developed CG-DIR algorithm was conducted on an NVIDIA Tesla C1060 card. This card is equipped with 240 1.3GHz stream processors and 4 GB graphic memory. Due to additional computational steps, the computational time is expected to be longer than the GPU-based original demons. In the six tested cases, the computational time for each CG-DIR iteration is about 1.3~1.6 times more than that of the original demons. For example, case HN 4 has an image size of 240×160×70, and it takes around 55 millisecond(msec) for each CG-DIR iteration compared to 42 msec for each demons iteration. For the cases examined in this study, total computational time ranges around 10~20 second.

## 4. Discussion

In image-guided adaptive radiotherapy, DIR is an essential tool for automated organ delineation, dose accumulation, and 4D treatment planning. Although this imaging tool works well with certain cases such as high-contrast thoracic 4D-CT images, it may generate DVFs with large uncertainties and errors in low-contrast regions leading to distortions in auto-propagated contours. Thus, manual contour inspection and editing is always necessary. The proposed CG-DIR algorithm incorporates the physician-edited contours into an additional deformable registration to reduce the errors in DVF and to improve the consistency between the DVF and the edited contours. This DVF correction is crucial for later dose accumulation in adaptive therapy or accurate 4D treatment planning.

    The contour-guided deformable image registration algorithm developed in this paper has an underlying assumption that contours edited by physicians are accurate. Intra-observer contouring variations exist in real clinical situations, which could lead to uncertainties in contours correction and subsequently DVF correction. However, the application of the CG-DIR algorithm in adaptive radiotherapy incorporates auto-propagated contours edited by physicians. Edited contours have much less variation than those independently drawn contours from scratch (Wang *et al.*, 2008). Hence, intra-observer variation will unlikely be a significant issue for the CG-DIR application.

    The computational efficiency of CG-DIR has room to be further improved. Under the scenario that physicians will edit contours based on auto-propagated contours, the registration in coarse-level with demons can be eliminated. The DVF generated during the auto-contouring can be used as an initial DVF value for CG-DIR and thus eliminating the need for coarse level registration. Regarding the memory usage of CG-DIR, in this study, each delineated structure has its own modified image, which is a memory-consuming strategy. The usage of GPU-based CG-DIR algorithm is limited by the available memory space on a given graphic card. As mentioned above, compared to demons algorithm, CG-DIR algorithm requires extra memory to store auxiliary variables. On a Telsa C1060 card, 4GB graphic memory allows us to handle images with a size ~256×256×70 and two delineated organs. Further investigation of optimizing memory usage is needed for applying this algorithm to CT data sets of larger size.

    Another issue in the CG-DIR algorithm is the selection of a proper regularization parameter λ. In this study, we use a trial-and-error method to find a good λ value for each study case. According to our observations, $\lambda_i$ values are relatively consistent for all studied cases. Additional comprehensive image datasets for other anatomical regions are needed to further confirm the findings presented here.

**5. Conclusions**

A contour-guided deformable image registration algorithm has been developed and validated for adaptive radiotherapy. Physician edited contours in conjunction with images intensities are employed to deform images more accurately, resulting in DVFs that are more consistent with physician-edited contours. This CG-DIR algorithm has been applied to patients, resulting in improved DVF accuracy with a high computational efficiency.

**Acknowledgements**

This work is supported in part by the University of California Lab Fees Research Program and by an NIH/NCI grant 2F32 CA154045-01.